\documentclass[a4paper,11pt]{article}
\usepackage{xy}

\begin{document}

\begin{center}{ QCD monopoles, abelian projections and gauge invariance}\end{center}
\vspace{0.3cm}
\begin{center}
A. Di Giacomo
\end{center}
\begin{center}
 Pisa University and INFN Sezione di Pisa
\end{center}
\vspace{2cm}

\begin{abstract}
It is shown that the creation of a monopole is a gauge invariant process, based on topology. Creating a monopole is independent on the abelian projection
in which it is created.  This is fundamental in defining an order parameter for detecting dual superconductivity of the QCD vacuum.
\end{abstract}

\section{Introduction.} \label{section1}
Quarks have never been observed as free particles, neither in ordinary matter, nor as  products of high energy collisions, in spite of the extensive experimental searches performed during the years, starting soon after  they were proposed as fundamental constituents of hadrons \cite{GM}\cite{ZW}, and long before $QCD$ were established as the theory of strong interactions. The result are upper limits on the abundance $n_q$ of quarks in nature, and on the inclusive cross section $\sigma_q$ for production of free quarks, e.g. in proton-proton collisions\cite{PDG}:
\begin{equation}
R_n \equiv \frac{n_q}{n_p} \le 10^{-27}  \hspace{2cm} R_{\sigma} \equiv \frac{\sigma_q}{\sigma_{TOT}} \le 10^{-15} \label{exp}
\end{equation}
 In Eq(\ref{exp})  $n_p$ is the abundance of protons, $\sigma_{TOT}$ the total cross-section. 
 
 In $QCD$, if quarks could exist as free particles,
 the expectation would be   \cite{okun}
 
 \vspace{0.5cm} 
 
\hspace{1.8cm} $R_ n \approx 10^{-12} $  \hspace{2.5cm} $ R_{\sigma} \approx 1$
 
 \vspace{.5cm}
 
 The existence of free quarks is  inhibited, at least by a factor  $10^{-15}$.
 This phenomenon is called colour confinement, or
 simply confinement.  $QCD$, which is to a very high degree of reliability the theory of strong interactions, must have some built-in mechanism
 at large distances which produces  confinement. $QCD$ at large distances is not understood analytically and can only be studied with non perturbative tools like e.g. Lattice. 
 
 In any case the only natural explanation of such an extreme inhibition factor is that  $R_n$ and $R_{\sigma}$ are  strictly zero, protected by some symmetry, i.e. that confinement is an absolute property,  deconfinement  is a change of symmetry and hence a true phase transition with an order parameter.
 
 This request is met by the mechanism of dual superconductivity of the $QCD$ vacuum first proposed in Ref\cite{'tHP} \cite{m}. The idea is that monopoles condense in the ground state of the confining vacuum in the same way as Cooper pairs do in the ground state of ordinary  superconductors, thus producing, by (dual) Meissner effect, chromo-electric flux tubes between quark- antiquark pairs and with them confinement.
 
 The idea is attractive a-priori. The symmetry in question can not be a subgroup of the gauge group, since the order parameter is forced to be gauge invariant \cite{Elitz}. As we shall see in detail below,  possible gauge invariant degrees of freedom live on the surface $S_2$ at spatial infinity: this means topology and the natural topological excitations in 3 spatial dimensions are magnetic charges (monopoles).
 
 Monopoles do exist in non-abelian gauge theories as classical static solutions of the field equations (solitons). This was first shown  in Ref.'s \cite{'tH1} \cite{Pol} in a $SU(2)$  Higgs model with the Higgs field in the adjoint representation. The role of the Higgs field is to produce, in the Higgs broken phase, a non trivial mapping of the sphere $S_2$ at spatial infinity on the gauge group. The magnetic charge is the corresponding winding number. In the unbroken phase there is no field on $S_2$ and no monopole. The magnetic charge  is  gauge invariant: indeed it is defined in terms of the Higgs field on $S_2$, which is invariant under ordinary gauge transformations [Sect \ref{Higgs}  below].  Formally a generic gauge transformation $U_G(x)$ can be written as a product
 \begin{equation}
 U_G(x) = U(x) U_B(x) \label{gt}
 \end{equation}
 with $U(x)$ the usual gauge transformation in the bulk of the system, which tends to the unit transformation on the border, and $U_B(x) =1$ at finite $x$ and non trivial on the border. If the fields vanish at infinity the theory is blind to $U_B$ and $U(x)$ is the gauge transformation. 
 If there is spontaneous breaking the Higgs field is non zero on the border and sensitive to $U_B$: however its value on the border is invariant under the transformations $U(x)$ which act on the quantum fluctuations.
 
 The component of the monopole solution along the Higgs field is a Dirac monopole. In addition the Higgs field has a zero, which is conventionally assumed to be the position of the monopole. We shall briefly recall these properties in Sect.\ref{Higgs}, to set the notation and the basic concepts. We shall do that for gauge group $SU(2)$. The extension to arbitrary compact groups will be presented in Sect.\ref{ext}.
 In Sect.\ref{Higgs} we shall also demonstrate that the existence of a monopole is a gauge-invariant concept.
 
 \vspace{.3cm}
 
In $QCD$ there is no fundamental Higgs field to produce a spontaneous breaking and with it a non trivial mapping of $S_2$ on the gauge group. However operators transforming in the adjoint representation can act in principle as effective Higgs fields. Monopoles are expected to sit on the zeroes of such operators \cite{'tH}. In fact a zero of an operator in the adjoint representation is a necessary and not generally sufficient condition for the existence of a monopole. We shall discuss this point in detail in Sect. \ref{QCD}     and show that for  some choice of the operator monopoles do exist. 

In the literature the gauge in which an effective Higgs field  is diagonal is called an "Abelian Projection". Monopoles are Dirac monopoles in the $U(1)$ subgroup of the gauge group defined by the abelian projection. Again we are speaking of gauge group $SU(2)$: the extension to higher groups will be discussed in Sect.\ref{ext}.

On the lattice monopoles are detected as $U(1)$ monopoles by the technique of Ref.\cite{dt}, which consists in observing the extra flux carried by the Dirac string. It happens  that generically both the number and the location of the observed monopoles strongly depend on the choice of the abelian projection. It is commonly  accepted that monopoles belonging to different abelian projections are different. It is not understood to what extent they are independent on each other. However monopoles are topological in nature, their existence is a gauge invariant concept, and cannot depend on the choice of the gauge. We shall discuss a  possible way  to solve this contradiction  in Sect. \ref{disc}.

 The monopoles which are supposed to condense to produce confinement are believed to belong to a specific projection. The most popular choice is the so called Maximal Abelian Projection \cite{'tH}, the motivation being that they show a better "Monopole Dominance"  \cite{suz} \cite{pol}. These are positive  credentials. However no convincing argument exists relating monopole dominance to monopole condensation. Attempts to determine numerically  the effective potential for monopoles from the lattice configurations and to show that it has a negative quadratic term responsible for condensation never worked \cite{pol}.

A more direct approach consists in defining
 the order parameter for monopole condensation as the vacuum expectation value of a  magnetically charged operator \cite{Dig}.

 An order parameter was defined and developed in Ref.'s \cite{Dig} \cite{DP}\cite{ddpp} \cite{dlmp} \cite{ bcdd}. The approach was successfully tested in the simple case of $U(1)$ lattice gauge theory \cite{DP}.

To extend the construction to non abelian theories, specifically to $QCD$, a key problem was the identification of the monopoles that should condense in the vacuum to produce dual superconductivity, i.e. of  the appropriate abelian projection.
The strategy was then  to define the order parameter as the vacuum expectation value of an operator  creating a monopole in  the $U(1)$ subgroup diagonal in some abelian projection and test its relation to confinement. The test can be done by comparison to an independent criterion  of confinement such as the force between static quarks, and this is possible at least in pure gauge theories. 

The most natural choice would have been the Max Abelian gauge, which was extensively studied in other approaches to the problem. However this choice was extremely demanding from the computational point of view, the Max abelian gauge being defined by an optimisation procedure, to be operated on each configuration in the numerical updating procedure.  An alternative possibility was to average on the choice of the abelian projection, assuming that they are somehow all equivalent \cite{'tH}: the monopole was created for $SU(2)$ in the conventional 3 direction used to update the configurations \cite{ddpp} \cite{dlmp}, which is randomly distributed with respect to physical directions in colour space.

 In this paper we shall prove  that this is indeed the correct procedure. When creating a monopole in a given abelian projection a monopole is created in all abelian projections. It could well be that some choice of the monopoles is more convenient than others for some purpose, but, at least to define the order parameter, all abelian projections are equivalent. Creation of a monopole and monopole condensation are abelian projection independent facts.
 
 In the initial stages of the program our order parameter seemed to work correctly for pure gauge theories with gauge group $SU(2)$ and $SU(3)$ \cite{ddpp} \cite{dlmp}. The order parameter $\langle \mu \rangle$ was $\neq 0$ in the confined phase and  zero in the large volume limit of
  the deconfined one. 
  
  In trying the same procedure for the exceptional group $G2$ infrared problems appeared: the order parameter was zero in the large volume limit both in the confined and in the deconfined phase\cite{cossu}.  The origin of the problem was understood by use of a strong coupling expansion \cite{bcdd} and proved to be independent on the choice of the gauge group.  The same disease was also present but numerically less visible in the case of $SU(2)$ and $SU(3)$. It  was small within errors to be observed  in the relatively small lattices and statistic used in \cite{ddpp} and \cite{dlmp}, but it showed up at larger lattices and statistics. An improved order parameter was constructed, with the same symmetry properties but with no infrared problem \cite{bcdd} , which is non zero in the confined phase, vanishes at large volumes in the deconfined phase, and obeys the scaling properties required by the class of the deconfining transition. A detailed numerical   test was done for $SU(2)$ \cite{bcdd},
 and is under way for $SU(3)$ and $G2$ gauge groups.
 
 Also for the improved version the monopole is created in an arbitrary abelian projection, and this is correct as we show below. The net result is that 
 we have now all the tools to disprove the mechanism of confinement by dual superconductivity or to find evidence for it.

 \section{Monopoles in gauge theories: the $SU(2)$ Higgs\\ model.} \label{Higgs}
 Monopoles can exist as end points of abelian magnetic flux lines in non abelian field theories. This was first shown in Ref.'s  \cite{'tH1} \cite{Pol} where an explicit magnetically charged static classical solution 
 was found of the field equations in the spontaneously broken phase of a SU(2) Higgs model with the Higgs field $\vec \Phi$ in the adjoint representation. 
 
 In the usual formulation of field theory  fields  vanish and local gauge transformations  become trivial at infinity, in particular on the sphere $S_2$ at spatial infinity for static solutions.
 In the broken phase of the Higgs model the Higgs field has non zero value and constant length in all points of  $S_2$. If we denote by  $G$ the gauge group a mapping exists of $S_2$
 on $G/U(1)$. If the group $G$ is $O(3)$   [$SU(2)$ ]  this mapping can be non trivial, i.e.  non connectable to the identity by continuous transformations. A winding number, $M$, is attached
 to any classical solution, a positive or negative integer or zero
  which is the magnetic charge  in units of $\frac{1}{2g}$, ($g$ $\equiv$ the gauge coupling). 
  Topology makes the solutions stable with respect to quantum fluctuations which vanish on $S_2$, and is itself invariant under the usual local gauge transformations acting on the fluctuations, which are trivial at infinity. The existence of a  monopole is a gauge invariant concept. Saying that $SU(2)$ breaks to $U(1)$ does not mean a breaking of gauge invariance, but only of the global $SU(2)$ on $S_2$.
  
  To be definite let us concentrate on $SU(2)$ gauge group where the Lagrangian is, in the usual notations,
 \begin{equation}
 L = -\frac{1}{4} \vec G_{\mu \nu} \vec  G_{\mu \nu} + \frac{1}{2} D_{\mu}\vec  \Phi D_{\mu}\vec  \Phi  -\frac{m^2}{2} \vec \Phi \vec \Phi  - \frac{\lambda}{4} (\vec \Phi \vec \Phi)^2 \label{lagr}
 \end{equation}
 The monopole solution is constructed \cite{'tH1}  \cite{Pol} in the so called "Hedgehog Gauge " in which the orientation of the field $\vec \Phi$ in colour space is the same as that of the position vector $\vec x$ in physical space. Explicitly, in the notation of Ref\cite{Shnir}
  \begin{equation}
 \Phi_a = \frac{x_a }{gx^2} H(\xi) ,   \hspace{3.0cm}    H(\xi) _{\xi \to \infty}\approx \xi,    \hspace{.5cm} \lim_{\xi \to 0} \frac{H(\xi)}{\xi}  =0      \label{eq1}
 \end{equation}
 \begin{equation}
 A^a_0 =0 , \hspace{0.2cm}  A^a_i = - \frac{1}{g x^2} \epsilon_{i a b}x_b [1 - K(\xi)]      \hspace{1.0cm} K(\infty) =0,\hspace{.4cm} K(0)= 1  \label{eq2}
 \end{equation}
 The conditions at the boundary come from the requirement that the energy be finite, i.e. that the monopole be a soliton.
 
  \vspace{.5cm}
  
  $\xi  = g \rho x$,  \hspace{.5cm} $ \rho \equiv \sqrt( -\frac{m^2}{\lambda}) $ .
  
 \vspace{.5cm}
 
 The equations of motion become, in terms of $H(\xi)$, $ K(\xi)$  \cite{Shnir}
 
 \begin{equation}
 \xi^2 \frac{d^2K}{d\xi^2} = K H^2 + K (K^2 -1) \label{eqK}
 \end{equation}
 
 and

 \begin{equation}
 \xi^2\frac{d^2H}{d \xi^2} = 2 K^2 H + \frac{\lambda}{g^2} H (H^2- \xi^2) \label{eqH}
 \end{equation}
 
  \vspace{.3cm}
  
 We have explicitly recalled Eq's (\ref{eqK}) (\ref{eqH}) since from them it easily follows by direct check that at small distances ($x \to 0$)   $H(\xi) \approx \alpha \xi^2 + O(\xi^3)$
 and $K(\xi) \approx 1 + \gamma \xi ^2 + O(\xi^3)$, with $\alpha$ and $\gamma$ numerical coefficients.  The gauge field is finite at small distances, vanishes at $\vec x =0$ and
 \begin{equation}
  \Phi_a \approx_{\vec x \to 0} constant \times x_a   \label{sd}
  \end{equation}
 The behaviour Eq (\ref{sd}) is a necessary condition for the solution to  be a monopole.
 
 Also the condition  Eq(\ref{eq1}) at infinity 
 \begin{equation}
 \Phi_a = \rho \frac{\hat x_a}{x} \label{ld}
 \end{equation} 
 
 with non zero $\rho$ is a necessary condition. It insures  a gauge invariant spontaneous breaking of the colour symmetry and the existence of a mapping 
 of $S_2$ on $SU(2)/U(1)$.  
 
 This argument is also supported by the computation of the magnetic charge  $M$ in terms of the functions $H$ and $K$. The result is
 \begin{equation}
  M =  \frac{1}{g} \int_0 ^{\infty} d\xi \frac{d}{d\xi} \left[\frac{H(1-K^2)}{\xi}\right] = \frac{1}{g}\left [\frac{H(1-K^2)}{\xi}\right]_0^{\infty}
 \end{equation}
Due to the behaviour of $H(\xi)$ and $K(\xi)$ at small $\xi$ the contribution at $\xi =0$ vanishes $O(\xi ^3)$. The contribution at large distances gives  the correct value $M= \frac{1}{g}$  iff  $\lim _{\xi \to \infty} \frac{H(\xi)}{\xi} =1$ or, by use of Eq (\ref{eq1})  $\Phi_a = x_a \rho  \neq 0$ on $S_2$.
 
 Eq (\ref{eq1}) describes the mapping $S_2 \to O(3)/U(1)$. The solution Eq (\ref{eq1}) for $\vec \Phi$ has a zero at  $\vec x=0$, which is conventionally the location of the monopole, and constant non zero length on $S_2$. 
 
 An alternative gauge is the unitary gauge in which the field $\Phi$ has a fixed orientation in the group, say $\hat n$, in all points of space: it is then constant on $S_2$, and has a singularity 
 at $\vec x =0$. In this gauge the solution for the gauge field  \cite{'tH1} \cite{Pol} looks at large distances like a point-like abelian magnetic charge in the  $U(1)$ subgroup which leaves the $\vec n$ axis invariant, plus a Dirac string or a Wu-Yang  2-dimensional  surface which carries the flux to infinity.  The Wu-Yang
  surface can be replaced by a straight string and the string can be rotated to any space direction, say the positive z axis since, because of the Dirac quantisation of the magnetic charge,  any string which crosses physical space is invisible.  
  
  The hedgehog gauge can be useful in locating monopoles as zeros of possible effective Higgs fields e.g. in $QCD$ \cite{'tH}.

   The unitary gauge privileges the description of the field on $S_2$.

   Modulo irrelevant quantum fluctuations a monopole is  identified by  the winding number of the mapping $S_2 \to O(3)/U(1)$ i.e. by the the flux in a string along the z axis of  the magnetic field parallel  to the Higgs field in colour space.
   
   In what follows we shall assume the solution of Ref. \cite{'tH1} \cite{Pol} as definition of monopole.

  Should there exist in the theory another scalar field of constant length in $S_2$, $\Phi'$,  transforming in the adjoint representation besides the Higgs field  $\Phi$ this would imply
  another mapping of $S_2$  on $O(3)/U(1)$ and again a possible existence of monopoles defined by the winding number of this mapping. The corresponding unitary representation would be called an abelian projection in the language  of Ref \cite{'tH}. $\Phi$ and $\Phi'$ define two different abelian projections. They transform in the same representation of $O(3)$ and therefore  a gauge transformation exists transforming them into each other on $S_2$. The transformation connecting the two unitary gauges is a global transformation.

  Consider  the quantity
    \begin{equation}
     M  =  \oint \hat \Phi(\vec x) \vec A_i(\vec x) dx_i  \label{mag}
     \end{equation}
     where $\hat \Phi$ is the orientation of the Higgs field $\vec \Phi$ in colour space and the integral is computed along a closed line in $S_2$,  specifically along a line which encircles the 
     singularity where the Dirac strings end.  By Stokes theorem, in the unitary gauge, where $\hat \Phi $ is independent on $\vec x$,
     \begin{equation}
     M = \int d  \sigma_i \hat \Phi \vec B_i 
     \end{equation}
     where now the integral is computed on a surface having the closed line as a border and $ \vec B_i = \epsilon_ {i j k} ( \partial_j \vec A_k - \partial _k \vec A _j)$.
      $M$  is nothing but the magnetic charge computed in the unitary gauge. In fact in a gauge in which $\ \hat \Phi$ is constant the 'tHooft tensor
      \begin{equation}
      F_{\mu \nu} \equiv \hat \Phi \vec G_{\mu \nu} - \frac{1}{g} \hat \Phi (D_{\mu} \hat \Phi  \wedge D_{\nu}\hat \Phi)
      \end{equation}
      reduces to
      \begin{equation}
        F_{\mu \nu} = \hat \Phi (\partial_{\mu} \vec A_{\nu} - \partial_{\nu}  \vec A_{\mu})
      \end{equation}
      The quantity $M$ of Eq.(\ref{mag}) is nothing but the flux across the string of the magnetic field defined by the 'tHooft tensor.

    Under a gauge transformation
      \begin{eqnarray}
        \hat \Phi & \longrightarrow & \hat \Phi'  =  R \hat \Phi  \\
        \vec A_{\mu} & \longrightarrow &   \vec A_{\mu}'  =R \vec A_{\mu}  + i \frac{1}{g} R^{\dagger} \partial _{\mu} R 
            \end{eqnarray}
      with $R$ is an $ \vec x$ dependent orthogonal matrix,  Eq(\ref{mag}) transforms to 
      \begin{equation}
      M' =  \oint \hat \Phi'(\vec x) \vec A'_i(\vec x) dx_i = M + i \frac{1}{g} \oint \hat \Phi(\vec x) \partial_i R^{\dagger}    dx_i    \label{mag1}
 \label{mag'}
      \end{equation}
      The second term vanishes if the gauge transformation is global [ $\partial_{i} R^{\dagger} =0$], as is the transformation between two abelian projections.
      
      The magnetic charge is abelian-projection independent.

It follows that if an operator $\mu$ creates a monopole \cite{ddpp} \cite{dlmp}\cite{bcdd} in one abelian projection, it creates a monopole in all abelian projections.

\section{Monopoles in gauge theories: QCD.}\label{QCD}
 
 In $QCD$ there is no Higgs field in the Lagrangian to allow monopole solutions. There could be, however, some field of the theory transforming in the adjoint representation which can act as an effective Higgs field thus allowing monopole solutions \cite{'tH}. We now analyse this possibility in some detail. 
 
 Again we discuss the case of the $SU(2)$ gauge group. The extension to higher groups will be discussed in Section 4.
 
 In formulae the statement that some operator in the adjoint representation, function of the gauge fields $\vec \Phi (\vec A_{\mu})$, exists which can act as an effective Higgs field
 is as follows.
 
 The effective Lagrangian is defined in terms of Feynman path integrals as 
 
 \vspace{.2cm}
 
 \begin{equation}
 \exp( -S_{eff}( \vec A_{\mu}, \vec \Phi) ) =  \delta( \vec \Phi - \vec \Phi(\vec A_{\mu}) )\exp(- S(\vec A_{\mu}) \label{Seff}
  \end{equation}
  
 \vspace{.3cm}
 
 To have monopole solutions $S_{eff}$ has to have the form Eq (\ref{lagr}). Moreover the two necessary conditions must be full-filled by the classical solution 
 
 \vspace{.4cm}
 
 1)  $\vec \Phi $ must have a zero and behave as Eq(\ref{sd}) around it.
 
 \vspace{.3cm}
 
  2) $\vec \Phi $ must have the form Eq(\ref{ld}) and be non zero on $S_2$.
  
  \vspace{.3cm}
  
  The condition 1) is satisfied at any zero of an operator transforming in the adjoint representation \cite{'tH}. Indeed, in the gauge in which $ \Phi \equiv \vec \Phi  \frac{\vec \sigma}{2}$ is diagonal (unitary gauge), assuming invariance under space rotations of the solution,  we have in the vicinity of $\vec x = 0$  
  $ \Phi = c r \frac{\sigma_3}{2}$ with $c$ a numerical constant. In the hedgehog gauge, this is nothing but  Eq(\ref{sd}). 
  
  Moreover if $\Phi$ is a reasonable function of $\vec A$ transforming in the adjoint representation it vanishes at $\vec x=0$ since $\vec A$ itself vanishes at $\vec x=0$ as from  Eq(\ref{eq1}) and from the behaviour of $K(\xi)$ at small values of $\xi$, $ K\approx 1+\gamma \xi^2 $.

  Things are more complicated with respect to the condition 2). It requires that the function $\vec \Phi (\vec A_{\mu}) $ entering in Eq(\ref{Seff}) be non zero and of constant length on $S_2$. This is impossible if  the function is a reasonable function of the fields  $\vec A_ i $, the space components of the vector potential, which vanish on $S_2$ by  the ansatz Eq(\ref{eq2}). 
  
  A possibility left is to relax the assumption $A^0 =0$ Eq.(\ref{eq2}) made in Ref.\cite{'tH1} and allow  $\Phi$ to depend on $\vec A_0$, which has to obey the equation  $\partial_0 \vec A_0 =0$ as required for a static  solution. The corresponding monopoles do exist and have been studied in the literature e.g. in Ref.'s \cite{JZ} \cite{DZ}
 \cite{Su}. We also refer to  \cite{dig6} where the formalism is developed for generic group and which will be useful in the following of this paper.

  The arguments above are  independent on the presence of matter fields like quarks in the theory.
 
 In conclusion if we want to define a creation operator for monopoles to investigate monopole condensation as a mechanism for confinement, we can define it in any abelian projection
 since creation of a monopole is an abelian  projection independent concept. This justifies the procedure of Ref's \cite{ddpp} \cite{dlmp}\cite{bcdd} in which the monopole was created in the 
 subgroup $U(1)$ defined by the conventional 3-axis used in constructing the configurations on the lattice.
 
 The argument needs the existence of at least one effective Higgs field, i.e. a field which transforms in the adjoint representation and has constant length on the sphere $S_2$. One such operator is the gauge field $\vec A_4= i \vec A_0$ in the gauge $\partial_0 \vec A_0 =0$.

    \section{Extension to generic compact gauge groups.} \label{ext}

 The construction and the arguments of Section 1  can be extended to a generic compact gauge group. Monopoles can exist related to the mapping of $S_2$ on any $SU(2)$ subgroup of the gauge group. There exists an $SU(2)$ subgroup  for each positive root $\vec \alpha$  of the Lie algebra. In the standard notation if $E_{\pm \vec \alpha}$ are the generators corresponding to the roots $\pm \vec \alpha$ , $H_{i}   ( i=1,..r)$ the commuting generators and $r$ the rank of the group, the Lie algebra is
 
 \begin{equation}
[ H_i, H_j] =0    \hspace{2.4cm}    [ H_i , E_{\pm \vec \alpha} ] = \pm  \alpha _i E_{\pm \vec \alpha} 
 \end{equation}
\begin{equation}  
[ E_{\vec \alpha},  E_{\vec \beta} ]  = N_{\vec \alpha ,\vec \beta} E_{{\vec \alpha + \vec \beta} }     \hspace{1.2cm}                     
[ E_{ +\vec \alpha}, E_{-\vec \alpha} ] = \vec H  \vec \alpha
 \end{equation}
The subgroup $SU(2)$ attached to the root $\vec \alpha$ is defined in terms of the generators by a trivial renormalisation
\begin{equation}
T^{\alpha}_{\pm} = \sqrt{\frac{2}{(\vec \alpha  \vec \alpha)}} E_{\pm \vec \alpha}  \hspace{2cm}  T^{\alpha}_3 = \frac{\vec \alpha \vec H}{(\vec \alpha \vec \alpha)}
\end{equation}
      By this definition
      \begin{equation}
      [T^{\alpha}_3, T^{\alpha}_{\pm}] = \pm T^{\alpha}_{\pm} \hspace{1.5cm} [T^{\alpha}_{+}, T^{\alpha}_{-} ] = 2 T^{\alpha}_3 \label {su2s}
      \end{equation}
      
      We shall first consider monopoles corresponding to the simple roots of the algebra. There are $r$ of them, with $r$ the rank of the group. The monopole solutions    in the associated $SU(2)$ subgroups are, in the  hedgehog gauge,
      \begin{equation}
      \Phi (\vec x)^{i }= \rho^{i} [\chi ^{i,a} ( \vec x ) T^{i}_a  + \mu^{i} - T^{i}_3 ]    \hspace{1cm}  \chi ^{i,a}(\vec x) = \frac{x^a}{x} \chi ( \xi ^{i}) \hspace{1cm} \chi(\infty) =1 \label{eqq1}
      \end{equation}
      with $\rho^i$ a constant, the length of $\Phi^i $ on $S_2$, and 
      \begin{equation}
       A^{(i)}_k = A^{(i)}_{k a} (x) T^{(i)}_a  \hspace{2cm}     A^{(i) }_{k a} (x) =  -\frac{1}{g} \epsilon_{a k j} \frac{x^j}{x^2}(1- K(\xi^{(i)})) \label{eqq2}
      \end{equation}
      The notation in Eq(\ref{eqq2}) is the same as in Eq(\ref{eq2}). In Eq(\ref{eqq1}) for convenience we use instead of $H(\xi)$ of Eq(\ref{eq1}) $\chi (\xi)= \xi H(\xi)$.

      The index $i$ runs over the simple roots.  $\mu^i$ is the fundamental weight  corresponding to the simple root  $i$.
      
      \vspace{.5cm}
      
      $\mu^i = \vec c^i . \vec H \hspace{2cm}$  $[\mu^i, T^j_{\pm} ]= \pm c^i. \alpha^j T^j_{\pm} = \pm \delta_{ij} T^j_{\pm}$
      
      \vspace{.5cm}
      
       A gauge rotation in  the $SU(2)$ subgroup   Eq(\ref{su2s}) leaves  $\mu^i - T^i_3 $  Eq(\ref{eqq1}) unchanged and can  transform   $ \chi ^{i,a} ( \vec x ) T^i_a$ to $\chi ( |\vec x |) T^i_3$ which is the unitary representation. 
       
        In this representation the Higgs field  is a constant on $S_2$ $\hspace{.2cm}\Phi^i =\rho^i \mu^i  \hspace{.1cm}$, so that the 'tHooft tensor $F_{\mu \nu}$  for whatever group \cite{dlp} reduces to  
       \begin{equation}Tr [ \mu^i (\partial _{\mu} A^i_{\nu} - \partial _{\nu} A^i_{\mu} )]
       \end{equation}  
       
       All the other components of the gauge field strength belonging to $SU(2)$ subgroups corresponding to simple roots other than $i$ do not contribute since their diagonal parts
      have $\vec c^j$   orthogonal to  $\vec c^i$ ,$  \hspace{.2cm} j \neq i$.
      
      Like in the $SU(2)$ case the solution appears in the unitary gauge as a point-like monopole in the colour direction projected by $\mu^i$ plus a Dirac string ending on $S_2$.
      A different abelian projection would lead to a similar result up to a redefinition of the $SU(2)$ direction in which the generator is  diagonal.
      As in the $SU(2)$ case the two unitary representations would differ by a global $SU(2)$ transformation  and the configurations would have the same topology.
      Again creating a monopole  \cite{ddpp}\cite{dlmp}\cite{bcdd}  is an abelian projection independent operation.
      In the case of gauge group $SU(N)$ simple roots are permuted by Weyl transformations\cite{HUMP}. Weyl reflections are global gauge transformations, and therefore the scales $\rho_i$ are all equal by symmetry to a single scale $\rho$. More symmetries exist as discussed
      in the specific model \cite{dig6} of Sect.{\ref{Section 5}.
      Similar analysis can be done for the monopoles associated to the subgroups $SU(2)$ different from those related to simple roots with minor complications.
      
      However they are more massive and less likely to condense in the vacuum than those associated to simple roots at least in a model in which $A_4$ acts as a Higgs field \cite{dig6} [See Section  3.]
      \section{ Monopoles and the Polyakov line.} \label{Section 5}
      
      In this section we shall elaborate on the results of Ref \cite{dig6} and assume that  the effective Higgs field which defines the monopoles condensing in $QCD$  confining vacuum  is  the euclidean time component $A_4= i A_0$ of the gauge field  \cite{JZ}. As usual $x_4 \equiv i x_0$ In the gauge $\partial_4 A_4 =0$  $A_4= A_4(\vec x) $ and the Polyakov line 
      \begin{equation}
      L(\vec x) = P\exp (\int^{\frac{1}{T}} _0 ig A_{4} (\vec x, \tau) d\tau)
      \end{equation}
      simplifies to
      \begin{equation}
      L (\vec x) = \exp(\frac{ig A_4(\vec x)}{T}) \label{polloop}
      \end{equation}
       $T$ is the temperature.  The gauge in which $A_{4}(\vec x) $ is diagonal coincides with the gauge in which the Polyakov line is diagonal. 
       
       In the unitary representation $A_4$ is constant on $S_2$ and all the monopole solutions must have it as boundary condition of the effective Higgs field. This is consistent 
         with the  general theorem \cite{gpy} that for a soliton solution like a monopole the quantity $\frac{1}{d}  Tr [ L( \vec x)]$ has a limit as $\vec x \to \infty$ independent 
        on the direction,  $\frac{1}{d}  Tr [ L( \infty)]$ , known as holonomy. 
    
       In addition, together with the invariance under gauge group transformations, this fixes the scale length of all the monopoles and their masses. In particular for $SU(N)$ gauge group
       the monopoles corresponding to the simple roots have  the lowest mass.  All the others have mass  multiple of the lowest mass by integers $> 1$  \cite{dig6}.

        The quantity 
        \begin{equation}
         \langle L \rangle = \frac{1}{V} \int d^3x \frac{1}{d}  Tr [ L( \vec x)]  \label{OP}
        \end{equation}
     is the order parameter for confinement in the absence of quarks (quenched $QCD$ ): $ \langle L \rangle =0$ in the confined phase,   $\langle L \rangle =1$ in the deconfined phase. It is easy to prove that 
     \begin{equation}
      \langle L \rangle = \frac{1}{d}  Tr [ L( \infty)] \label{hol}
     \end{equation}
     $d$ is the dimension of the representation.
     Eq(\ref{hol}) relates the order parameter for confinement to the value on $S_2$
     of the effective Higgs field  in the unitary gauge, $A_4(\infty)$.
     
     For a generic gauge group   $ A_4( \infty)  = \bar \rho C $ with $C = \sum_{1}^{r} \mu_i$,  $\mu_i$ the $i-th$ fundamental weight  and $\bar \rho$ a constant with dimension of a mass, which is the inverse size of the lightest monopole \cite{dig6}. For $SU(N)$ gauge group the result is, putting $\alpha =\frac{ \bar \rho}{T} $,
     \begin{equation}
     \langle L \rangle = \frac{1}{N} Tr [exp(i C \alpha)]= \frac{1}{N} \frac{\sin(\frac{\alpha N}{2})}{\sin(\frac{\alpha}{2})}
     \end{equation}
     There is deconfinement ($\langle L \rangle  =1$ ) when there are no monopoles or $\bar \rho =0$. Confinement requires $\langle L \rangle  =0$ or $\alpha = \frac{2 \pi}{N}$.
     For $N\ge 3$ the transition is first order and the order parameter has a discontinuity, for $N=2$ it is second order and the change in $\alpha$  is expected to be continuous.
     
     At temperatures smaller than the inverse correlation length $\Lambda$ of the system $\Lambda $ acts as an infrared regulator and $\frac{\bar \rho}{T}$ is replaced by $\frac{\bar \rho}{\Lambda}$ \cite{dig6}, so that the condition for confinement becomes $ \frac{\bar \rho}{\Lambda} =\frac{2 \pi}{N} $ : the size of the monopoles is directly
     related to the physical correlation length, and comparable with it for not very large $N$.
     
     Finally if one  thinks of the vacuum as a liquid or gas of monopoles, the distribution $P( \langle L \rangle) $ of the value of the  order parameter near its maximum $\langle L \rangle _{max}$
    only  gets contributions from the inner part of the monopoles: it depends on the form of the monopole effective Higgs field and, of course on the density of monopoles $\delta_M$ in the vacuum \cite{dig6}.  The exact form is 
    \begin{equation}
    P( \langle L \rangle ) = K \frac{4 \pi \delta_M}{3 \Lambda^3} ( \langle L \rangle _{max} - \langle L \rangle)^{\frac{1}{2}}
    \end{equation}
     with $K$ a constant depending on $N$ of the order of unity at moderate values of $N$ \cite{dig6}.
    Comparing with the value observed on the lattice e.g. for $SU(2)$ namely  $P(\langle L \rangle) = \frac{2}{\pi} ( 1-\langle L \rangle ^2)^{\frac{1}{2}} $ a density results  of the order of one monopole per sphere of radius $\frac{1}{\Lambda}$.  The natural scale for everything is the physical correlation length.
\section{Discussion} \label{disc}
 Monopoles do exist in the Hilbert space of gauge theories as classical configurations  with non trivial topology.  The total magnetic charge is Abelian Projection independent, and so is the operator which creates a monopole, and its vacuum expectation value which is the order parameter for monopole condensation.
 
  As a consequence of the results of this paper the creation operator of a monopole is perfectly defined and the procedure of Ref. \cite{dlmp} in the improved version of Ref \cite{bcdd} well established. The vacuum expectation value of the operator which creates a monopole can be computed numerically. If it is non zero in the confined phase where monopoles condense and zero in the deconfined phase, dual superconductivity is confirmed as a mechanism for confinement. This is already true in pure gauge $SU(2)$ theory\cite{bcdd}. Study of $SU(3)$ and $G2$ gauge groups is in progress.
   
  The  role of the Polyakov line as effective Higgs field is not fully understood and deserves further study.
  
  A final comment on the monopole phenomenology on the lattice. Monopoles  are detected on the lattice  by use of the procedure of Ref \cite{dt} applied to the subgroup $U(1)$ identified by the abelian projection. Looking e.g. to the plaquettes  at the boundary of a  spatial elementary cube any value of the magnetic flux  $\theta$ through a plaquette which is out of the range $-\pi \le \theta \le  \pi$ is interpreted as a Dirac string which signals the presence of a monopole (or antimonopole) inside the cube. If this is true the monopole should stay in the same cube by changing abelian projection e.g. by a time independent gauge transformation. This is generally not the case. Similarly in updating the configuration the monopole should stay for many steps as a quasi stable background classical configuration, as happens for the instantons, and again this is not generally the case.  All this is an effect  of the procedure used to detect monopoles which is plagued by lattice artefacts. The distinction traditionally present in the literature between "short" and "long" monopoles is a reflection of that as well as the attempts to get rid of the artefacts in more recent papers \cite{SUZ}.  The presence of artefacts can be understood as follows.  The magnetic flux of a monopole is distributed on the surface of a sphere of surface  $S$ centered on it in such a way that each plaquette has a small fraction of it if the sphere is large enough: if in addition fluctuations are small (as is at high $\beta$ )  the existence of a Dirac string would be detected unambiguosly as a plaquette with large magnetic flux  by the method of Ref\cite{dt}. If instead the volume considered is small, e.g. an elementary cube and the flux through a plaquette of the order of $\pi$ lattice monopoles can show up as local fluctuations with no topological content. To avoid that  one should work in a situation in which the statistical noise is negligible. This can be done by going at high $\beta$ and large lattices or using such well known techniques like smearing or cooling, which depress short range fluctuations and expose configurations which are stable because of topology.  All of this deserves further study.  
  
  In conclusion a physical monopole is defined by its topology. Monopoles as defined on the lattice may not be monopoles, but lattice artefacts.
\section{Aknowledgements}
 The author wishes to thank Claudio Bonati for discussions and for critical reading of the manuscript.

\end{document}